# Chapter XX

# Electromagnetic wave diffraction by periodic planar metamaterials with nonlinear constituents

**V. Khardikov[1,2], P. Mladyonov[1], S. Prosvirnin[1,2], V. Tuz[1,2]**

[1]Institute of Radio Astronomy of National Academy of Sciences of Ukraine,
4, Krasnoznamennaya st., Kharkiv 61002, Ukraine,

[2]School of Radio Physics, Karazin Kharkiv National University,
4, Svobody Square, Kharkiv 61077, Ukraine

**Abstract** We present a theory which explains how to achieve an enhancement of nonlinear effects in a thin layer of nonlinear medium by involving a planar periodic structure specially designed to bear a trapped-mode resonant regime. In particular, the possibility of a nonlinear thin metamaterial to produce the bistable response at a relatively low input intensity due to a large quality factor of the trapped-mode resonance is shown. Also a simple design of an all-dielectric low-loss silicon-based planar metamaterial which can provide an extremely sharp resonant reflection and transmission is proposed. The designed metamaterial is envisioned for aggregating with a pumped active medium to achieve an enhancement of quantum dots luminescence and to produce an all-dielectric analog of a 'lasing spaser'.

## 1.1 Introduction

One of the current trends in the theory of metamaterials is the development of two-dimensional planar periodic systems (metasurfaces, metamaterials) constructed in the form of arrays of resonant metallic or dielectric particles, which are arranged periodically on a thin (compared with the wavelength) dielectric layer. It is known that such planar metamaterials can create an environment whose electromagnetic characteristics are similar to those achieved in the traditional cavity resonators, but, unlike the latter, planar structures can have a much smaller size.

The metamaterial optical properties significantly depend on the resonant features of its constituent particles. It turns out that the particles with a special form of symmetric split rings or squares exhibit resonant properties, which result in a sudden change in the effective parameters of the metamaterial in a certain frequency band [1, 2]. Such resonances have quasi-static nature, since the size of the unit cell of the metamaterial is small. Thus, the particles can be seen as an oscillatory circuit, which has its eigen frequency and quality factor. Unfortunately, the presence of large radiation losses which appear due to the strong electromagnetic coupling of the system to free space, and a relatively small size (compared with the wavelength) of particles do not allow to reach the high-$Q$ resonances in conventional planar metamaterials.

Nevertheless there is a possibility to achieve strong electromagnetic field confinement and localization in planar metamaterials which support a trapped-mode resonant regime [3, 4]. These resonances exist in two-dimensional planar periodic metamaterials with complex doubly or multi-connected metallic or dielectric particles, which have a low degree of asymmetry. In the near-infrared band it was shown theoretically [5] and confirmed experimentally [6] that introducing two slightly asymmetric metallic elements into the periodic cell can lead to the anti-phase current trapped-mode excitation. In this case the electromagnetic coupling of conductive elements with free space is very weak, which provides low radiation losses and, therefore, high $Q$-factor resonances.

Such strong field localization in the mentioned metamaterials opens prospects for their application in laser and nonlinear optics. Thus, in [7, 8], the idea of using resonant enhancement of the electromagnetic field which is strongly localized on the surface of metallic nanoparticles is proposed to create nanoscale devices, in order to amplify or generate radiation in the visible and infrared bands. Further, in the development of this idea, a compact planar periodic structure is considered [9]. The proposed system provides strong field localization due to the trapped-mode excitation and acts like a conventional laser cavity.

In the present days, the theory of nonlinear metamaterials is actively developing [10-13] in the fields of controlling light with light [14], and parametric conversion of optical harmonics [15]. Here a particular interest is to study the peculiarities of intense light interaction with planar structures which sustain the trapped-mode resonant regime, and contain nonlinear components. In our opinion, strong field localization, which can be achieved in such structures, opens wide prospects for their application in the area of nonlinear optics.

## 1.2 Planar Metamaterials with Metallic Particles

### 1.2.1 Trapped-Modes: Concept

Let us assume an electromagnetic plane wave

$$\vec{E}^i = \vec{P}\exp[-j(\vec{k}^i \cdot \vec{r})] \tag{1}$$

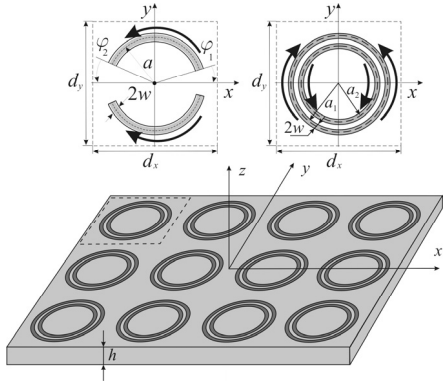

**Fig. 1.** Fragment of the planar metamaterial and its elementary unit cells.

incidents on a doubly periodic planar array of identical particles with a complex shape which are placed on a thin dielectric substrate. In (1) $\vec{P}$ is the polarization vector. Throughout this chapter the time dependence is assumed to be $\exp(j\omega t)$. The reflected and transmitted fields can be represented as a superposition of partial waves

$$\vec{E}^r = \sum_{m=-\infty}^{\infty}\sum_{m=-\infty}^{\infty}\vec{a}_{mn}$$
$$\times \exp\left\{-j\left[\vec{\chi}_{mn}\vec{\rho} + \gamma(\vec{\chi}_{mn})z\right]\right\},\ z \geq 0, \tag{2}$$

$$\vec{E}^t = \sum_{m=-\infty}^{\infty}\sum_{m=-\infty}^{\infty}\vec{b}_{mn}\exp\left\{-j\left[\vec{\chi}_{mn}\vec{\rho} + \gamma(\vec{\chi}_{mn})(z+h)\right]\right\},\ z \leq -h, \tag{3}$$

where $\vec{\chi}_{mn} = \vec{e}_x(k_x^i + 2\pi n/d_x) + \vec{e}_y(k_y^i + 2\pi n/d_y)$, $d_x$ and $d_y$ are the periods of the array, $\vec{\rho} = \vec{e}_x x + \vec{e}_y y$, $\gamma(\vec{\chi}) = \sqrt{k^2 - \chi^2}$, and $\operatorname{Re}\gamma \geq 0$, $\operatorname{Im}\gamma \leq 0$.

Let us note that the propagation direction of any spatial partial wave in the reflected (2) and transmitted (3) fields depends only on directions of periodicity in the plane of array, sizes of its periods, the direction of initial plane wave incidence, and the wavelength. If a plane wave is incident on the double periodic array, the defined set of spatial partial waves is formed in space. As it can be easily derived, the propagation directions of spatial partial waves in this set do not change if the initial plane wave is incident upon the same array at any other direction provided that this direction coincides with the propagation direction of any spatial partial wave in the set.

The method of moments is generally used to solve the problem of electromagnetic scattering by arrays of metallic particles [16]. In the framework of this method it is implied that the metallic pattern is a very thin conductor. The method also takes into account the fact that the array is placed on a thin lossy dielectric substrate. By enforcing the impedance boundary condition

$$\vec{E}_{\tan}\Big|_{z=0} = Z_s \vec{J}_s,\tag{4}$$

a vector integral equation is derived which is related to the current induced on a particular particle (here $Z_s$ represents the surface impedance of this particular particle). The integral equation is reduced to an algebraic one by using the standard spectral-Galerkin technique. So, using the method of moments allows us to determine both the distribution and magnitude of the current $J$ which flows along the metallic particles and further to calculate the reflection $r$ and transmission $t$ coefficients.

As a result of calculations of the current distribution and optical response of planar metamaterials with particles of different shape it is revealed that if these particles possess specific structural asymmetry, in a certain frequency band, the antiphase current oscillations with almost the same amplitudes appear on the particles parts (arcs). The scattered electromagnetic field produced by such current oscillations is very weak, which drastically reduces its coupling to free space and therefore radiation losses. Indeed, both the electric and magnetic dipole radiations of currents oscillating in the arcs of the neighbor particles are cancelled. As a consequence, the strength of the induced current can reach very high value and therefore ensure a high-$Q$ resonant optical response. Such a resonant regime is referred to so-called 'trapped-modes', since this term is traditionally used in describing electromagnetic modes which are weakly coupled to free space.

The most remarkable property of the trapped-modes is that they allow in principle to achieve high-$Q$ resonances in a very thin structure. The trapped-modes are normally inaccessible in the systems with particles of a symmetrical form, but can be excited if these particles have a certain structural asymmetry that allows reaching weak coupling to free space. Nevertheless, in the arrays with symmetric configuration of particles the excitation of the trapped-modes is also possible if the shape of these particles is specially designed. It is important that in the latter case the system can become polarization-insensitive.

Further we consider two particular configurations of metallic particles with asymmetric [4] and symmetric [17] designs which can support the trapped-mode excitation (see Fig. 1). In the first case, the unit cell consists of metallic particles in the form of asymmetrically split rings (ASRs). Each ASR contains two identical strip elements positioned opposite each other. The right-hand split between the strips $\varphi_1$ is a little different from the left-hand one $\varphi_2$, so that the square unit cell is asymmetric with regard to the $y$-axis. In the second case, the unit cell of the studied metamaterial contains a single double-ring (DR). The radii of the outer and in-ner rings are fixed at $a_1$ and $a_2$, respectively. Suppose the width of the metal rings in both configurations is $2w$, and the arrays are placed on a thin dielectric substrate with thickness $h$ and permittivity $\varepsilon$. We consider a normal incidence on the structure of a linearly polarized monochromatic plane wave with an amplitude $A$ and frequency $\omega$. Assume that the direction of the vector $\vec{E}$ of the incident wave coincides with the direction of the $y$-axis.

In the case of such a polarization of the incident field, in the structure of the first type, the trapped-mode excitation can be reached. Due to the 4-fold symmetry of the unit cell of the structure of the second type, its resonant properties do not depend on the direction of the vector $\vec{E}$ of the normally incident wave, i.e., as it was mentioned before, the second configuration is polarization-insensitive.

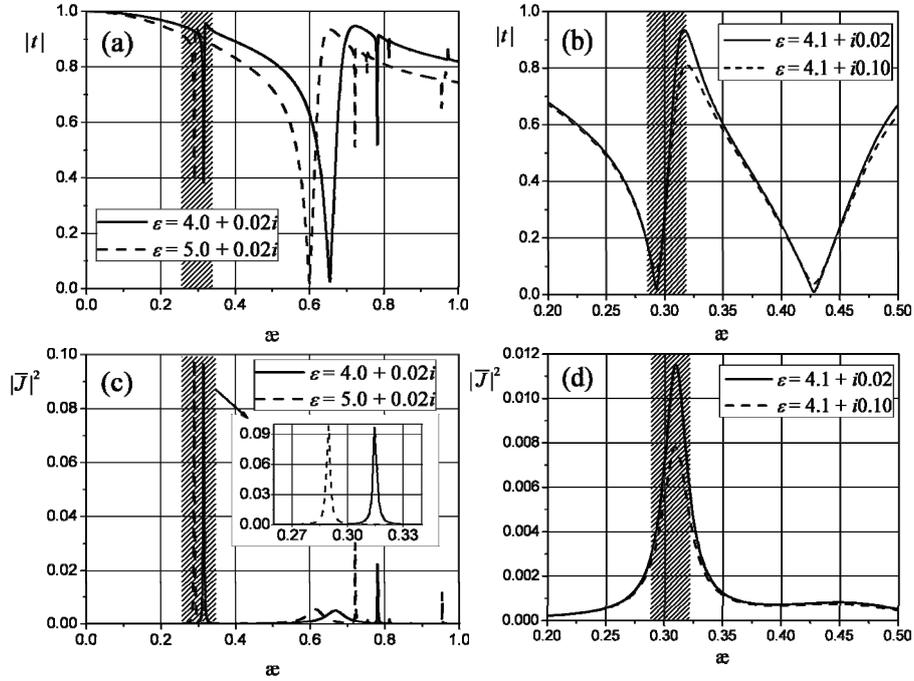

**Fig. 2.** Frequency dependences ($æ = d / \lambda$) of magnitudes of the transmission coefficient and squared average current (in a.u.) of the metamaterial with (**a**, **c**) ASR and (**b**, **d**) DR particles; (**a**, **c**) $d = d_x = d_y = 900$ nm, $\varphi_1 = 15^0$, $\varphi_2 = 25^0$; (**b**, **d**) $d = d_x = d_y = 800$ nm, .

Typical frequency dependences of the transmission coefficient and current magnitudes calculated with the method of moments for ASR and DR structures are presented in Fig. 2. One can see that at the dimensionless frequency nearly $æ = d / \lambda \approx 0.3$, a sharp resonance occurs in the structures of both types (see the shaded areas in Fig. 2). This resonance corresponds to the excitation of a trapped-mode because equal and opposite directed currents in the two arcs of each complex particle of array radiate a little in free space. The resonance has a high-$Q$ fac-tor, and the current magnitude reaches its maximum exactly at this frequency. Remarkably, as the permittivity of the substrate increases, the resonant frequency shifts down to low values (Fig. 2a). Also as the value of ohmic losses in the metamaterial substrate increases, the magnitude of current and quality factor of the resonance decrease (Fig. 2b) but, nevertheless, the resonance remains to be well observed. We expect that such a high-$Q$ resonant regime is promised to enhance some nonlinear effects, since at the frequency of trapped-mode excitation the field is strongly localized inside the system.

### 1.2.2 Inner Field Intensity Estimation

In order to understand the ability of the proposed structures to enhance some nonlinear effects, it is required to estimate the intensity of the inner field which is localized within the system. The special geometry of the metamaterial with symmetrical DR allows us to obtain an analytical expression for this demand [12, 13]. So, at the trapped-mode resonance, the electromagnetic energy is confined to a very small region between the rings. Therefore, the approximation based on the transmission line theory is used here to estimate the field intensity between the rings. According to this theory, conductive rings are considered as two wires with a distance $b$ between them. Along these wires the currents flow in opposite directions. Thus the electric field strength is defined as

$$E_{in} = V/b,$$
(5)

where $V = ZJ$ is the line voltage, $b = a_1 - a_2 - 2w$, $J$ is the magnitude of current which flows along the DR-element, and $Z$ is the impedance of line. The impedance is determined at the resonant dimensionless frequency $\ae_0 = d/\lambda_0$,

$$Z = \ae_0 \frac{60l}{dC_0},$$
(6)

where $l = \pi(a_1 + a_2)/2$, and

$$C_0 = \frac{1}{4} \ln \left[ \frac{p}{2w} + \sqrt{\left( \frac{p}{2w} \right)^2 - 1} \right],$$
(7)

is the capacity in free space per unit length of line, $p = a_1 - a_2$. From this model it follows that the electric field strength between the rings is directly proportional to the current magnitude $J$. Since the unit cell is small in comparison with the wave-length, the current magnitude $J$ can be substituted with its value averaged along the metallic ring, $\bar{J}$. From our estimations it can be concluded that the intensity of the incident field $I_{in}(\bar{J})$ which is enough for the nonlinearity to become apparent is about 10 kW/cm².

### 1.2.3 Optical Bistability and All-Optical Switching

The effect of optical bistability (or, generally, multistability) is a basis of numerous applications such as all-optical switching, differential amplification, unidirectional transmission, power limiting, pulse shaping, optical digital data processing, and others [14, 18]. A classical example of the bistable device is a Fabry-Perot interferometer filled with a Kerr-type nonlinear material. In this case, the resonator provides feedback, which is essential to obtain a multivalued intensity at the structure's output. However, in such a system, both relatively strong light power and/or large enough volume of the nonlinear optical material are generally needed to achieve a sizeable nonlinear response.

A promising way to realize an optical switching in compact devices can be found in using planar metamaterials. In particular, at the trapped-mode resonance the electromagnetic energy is confined to a limited extent around the particles, where the energy density reaches substantially high values. This makes the response of the metamaterial operating in the trapped-mode regime extremely sensitive to the dielectric properties of the substrate.

If a metamaterial is under an action of intense light (i.e. in the nonlinear regime), the substrate permittivity $\varepsilon$ becomes to be depended on the value of the average current $\bar{J}$ ($\varepsilon = \varepsilon_1 + \varepsilon_2 I_{in}(\bar{J})$). Thus, the appropriate average current magnitude for a given $\varepsilon$ can be found using the next nonlinear equation

$$\bar{J} = \tilde{A} F_{\bar{J}}(\omega, \varepsilon(I_{in}(\bar{J}))) , \qquad (8)$$

where $\tilde{A}$ is a dimensionless coefficient which depicts how many times the incident field magnitude $A$ is greater than 1 V/cm. Thus, the magnitude $A$ is a parameter of equation (8), and, at a fixed frequency $\omega$, the solution of this equation is the average current magnitude $\bar{J}$ which is depended on the magnitude of the incident field ($\bar{J} = \bar{J}(A)$). Further, on the basis of the current $\bar{J}(A)$ found by a numerical solution of equation (8), the permittivity of the nonlinear substrate $\varepsilon = \varepsilon(I_{in}(\bar{J}))$ is obtained and the reflection and transmission coefficients are calculated as functions of the frequency and magnitude of the incident field.

If the structure substrate is made of a Kerr-type nonlinear material, the curves of the average current magnitude versus incident field magnitude have a form of *S*-like hysteresis loops (Fig. 3a) [10]. Such a form of the input-output characteris-tic of the studied metamaterial is inherent to the most optical bistable devices. The presence of hysteresis results in abrupt switching between two distinct states with small and large levels of transmission nearly the frequency of the trapped-mode excitation (Fig. 3b).

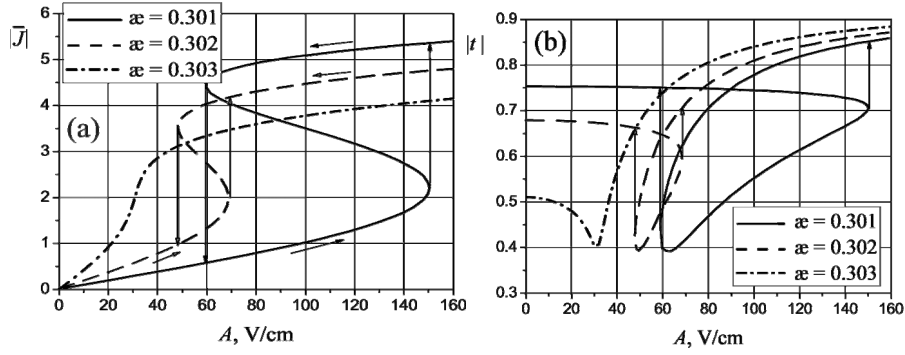

**Fig. 3.** The square of (**a**) the current magnitude (in a.u.) averaged along split ring and (**b**) the magnitude of the transmission coefficient versus the incident field magnitude in the case of the ASR nonlinear metamaterial ( $\varepsilon_1 = 4 + 0.02i$ , $\varepsilon_2 = 5 \times 10^{-3}$ cm$^2$/kW). The arrows indicate a bistable switching between two distinct levels of transmission. All other parameters are the same as in Fig. 2.

The origin of such a bistable response can be explained as follows. Suppose that the trapped-mode resonant frequency is slightly higher than the incident field frequency. As the intensity of the incident field rises, the magnitude of currents on the metallic particles increases. This leads to increasing the field strength inside the substrate and its permittivity as well. As a result, the frequency of the resonant mode decreases and shifts toward the frequency of incident wave, which, in turn, enhances further the coupling between the current modes and the inner field intensity in the nonlinear substrate. This positive feedback increases the slope of the rising edge of the transmission spectrum, as compared to the linear case. As the frequency extends beyond the resonant mode frequency, the inner field magnitude in the substrate decreases and the permittivity goes back towards its linear level, and this negative feedback keeps the resonant frequency close to the incident field frequency.

At once, the frequency dependence of the transmission coefficient magnitude manifests some impressive discontinuous switches to different values of transmission, as the frequency increases and decreases in the resonant range for the sufficiently large intensity of the incident wave. The shifting of the peak of the resonance and the onset of a bistable transmissivity through the ASR structure is similar to that of the reflection from a Fabry-Perot cavity (Fig. 4a). However, the trapped-mode resonance is Fano-shaped rather than the Lorentzian, as is the characteristic of 1D Fabry-Perot cavities. This Fano resonance can lead to a peculiar transmission spectra and bistable behavior. In particular, the transmission resonance of the ASR structure may loop back on themselves (Fig. 4b).

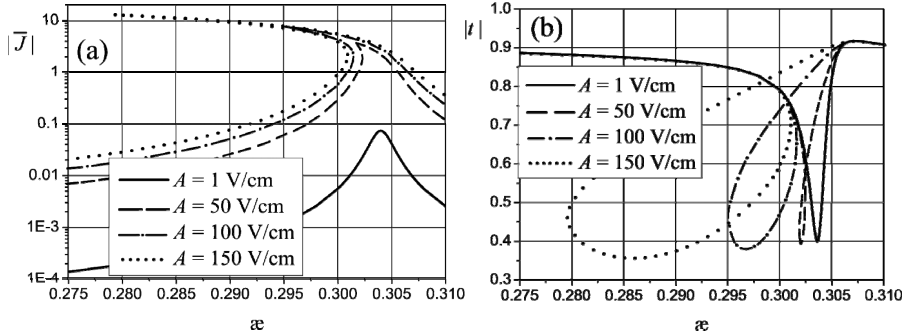

**Fig. 4.** Frequency dependences of (**a**) the square of the current magnitude (in a.u.) averaged along split ring (on the logarithmic scale) and (**b**) the magnitude of the transmission coefficient in the case of the ASR nonlinear metamaterial ( $\varepsilon_1 = 4 + 0.02i$ , $\varepsilon_2 = 5 \times 10^{-3}$ cm$^2$/kW). All other parameters are the same as in Fig. 2.

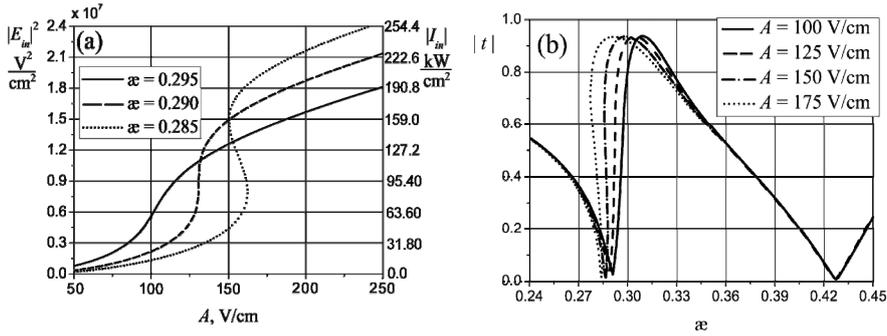

**Fig. 5.** Magnitude of the inner field intensity (**a**) versus magnitude of the incident field and (**b**) the frequency dependences of the magnitude of the transmission coefficient in the case of the DR nonlinear metamaterial ( $\varepsilon_1 = 4.1 + 0.02i$ , $\varepsilon_2 = 5 \times 10^{-3}$ cm$^2$/kW). All parameters are the same as in Fig. 2.

The most appropriate form for the realization of optical switching has the spectral characteristics of the DR-metamaterial [11-13]. While the dependence of the inner field intensity versus the incident field magnitude has the form of the *S*-like hysteresis loop (Fig. 5a), the frequency dependence of the magnitude of the transmission coefficient has a sharp asymmetric Fano-shape of the spectral line where the transmission coefficient changes from low to high level in a very narrow frequency range (Fig. 5b). Such a form of resonance is very suitable to obtain great amplitude of switching since there are gently sloping bands of the high reflection and transmission before and after the resonant frequency.

### 1.2.4 Strong Field Confinement in Bilayer-Fish-Scale System

Another type of metamaterials which supports Fano-shape trapped-mode resonances is a planar metamaterial which consists of an equidistant array of continuous meandering metallic strips placed on a thin dielectric substrate (the fish-scale structure [19]). In this system the trapped-mode resonance appears due to a special form of strips, and this form is designed in view of the polarization of the incident field. A way to expand the functionality of such a fish-scale structure lies in the placement of the second grating on the back side of a thin dielectric substrate. In this case additional trapped-mode resonance can appear due to interaction of the antiphase current oscillations between two adjacent gratings [20, 21]. This configuration is of particular interest in the case when the substrate is made of a field intensity dependent material (for example, a Kerr-type medium) because the strong field localization between the gratings can significantly enhance the nonlinear effects.

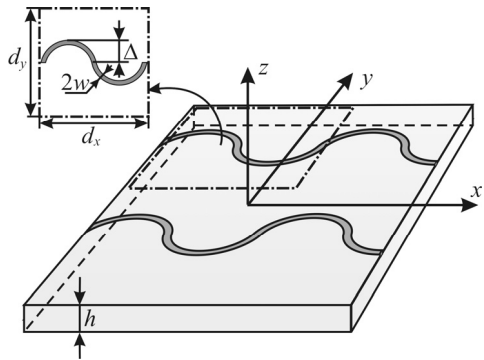

**Fig. 6.** Fragment of the bilayer-fish-scale metamaterial and its elementary unit cell.

We consider a bilayer structure which consists of two gratings of planar perfectly conducting infinite strips placed on each side ($z = 0$, $z = -h$) of a thin dielectric substrate (Fig. 6). The unit cell of the structure under study is a square with sides $d_x = d_y = d$. The width of the metal strips and their deviation from a straight line, respectively, are $2w/d = 0.05$ and $\Delta/d = 0.25$. Suppose that the normally incident field is a plane monochromatic wave polarized in parallel to the strips (*x*-polarization).

Due to the bilayer configuration of the structure under study there are two possible current distributions which correspond to the trapped-mode resonances. The first distribution is the antiphase current oscillations in the arcs of each grating. In this case the structure can be considered as a system of two coupled resonators which operate at the same frequency because the gratings are identical. We have labeled this resonant frequency in Fig. 7 by the letter $\text{æ}_1$. Obviously that the distance $h$ between the gratings will strongly effect on the resonant frequency position since this parameter defines the electromagnetic coupling degree. The *Q*-factor of this resonance is higher in the bilayer structure than that one existed in a single-layer structure but their similarity lies in the fact that the current magnitude in the metallic pattern depends relatively weakly on the thickness and permittivity of the substrate.

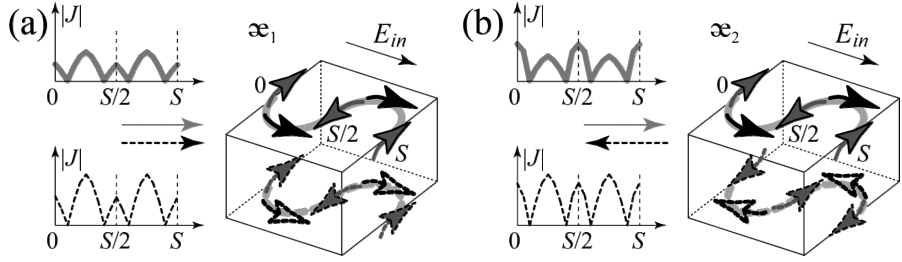

**Fig. 7.** The surface current distribution along the strips in the bilayer structure composed of gratings with wavy-line metal strips.

The second distribution is the antiphase current oscillations between two adjacent gratings. Similarly we have labeled this resonant frequency in Fig. 7 by the letter $æ_2$. It is well known that the closer are the interacting metallic elements, the higher is the $Q$-factor of the trapped-mode resonance. Thus varying both the distance between the gratings and substrate permittivity changes the trapped-mode resonant conditions and this changing manifests itself in the current magnitude. Remarkably that in this type of current distribution the field is localized between the gratings, i.e. directly in the substrate, which can sufficiently enhance the non-linear effects if the substrate is made of field intensity dependent material.

This circumstance is depicted in Fig. 8 where typical curves of the inner field intensity and the transmission coefficient magnitude are given as functions of the frequency and the incident field intensity in the nonlinear regime.

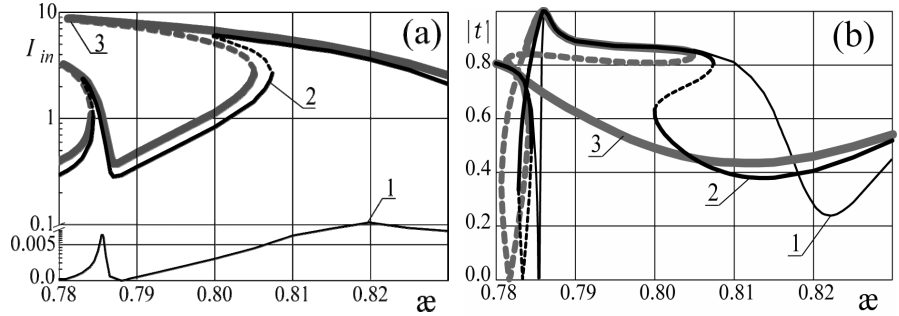

**Fig. 8.** The frequency dependences of (**a**) the inner field intensity (on the logarithmic scale) and (**b**) the transmission coefficient magnitude versus the different incident field intensity in the case of the nonlinear permittivity ( $\varepsilon = \varepsilon_1 + \varepsilon_2 |I_{in}|^2$, dimension of $I_{in}$ is in kW/cm²); $\varepsilon_1 = 3.0$, $\varepsilon_2 = 0.005$ cm²/kW; curve 1 — $I_0 = 1$ kW/cm², curve 2 — $I_0 = 200$ kW/cm², curve 3 — $I_0 = 300$ kW/cm².

For these calculations the structure parameters are chosen in such a way that the both resonances are closely settled and the frequency $æ_1$ of the first resonance is less than the frequency $æ_2$ of the second one ($æ_1 < æ_2$). One can see that as the intensity of the incident field rises, the frequency dependences of the inner intensity magni-tude take a form of the bent resonances and a bistable regime occurs. An important point is that this bending is different for the first and second type of resonances due to the difference in the current magnitude changing at these two frequencies. Thus at the resonant frequency $æ_2$ the magnitude of currents which flow along the strips of both gratings are significant, and they are greater than the magnitude of currents which flow nearly the resonant frequency $æ_1$ (Fig. 8a), and, in the nonlinear regime, the bending of the peak $æ_2$ is greater than that one of the peak $æ_1$. As a result, the spectral curves of the transmission coefficient magnitude experience different changes nearly the trapped-mode resonant frequencies. At the frequency $æ_1$ the curve transforms into a closed loop that is typical for the sharp nonlinear Fano-shape resonances. The second resonance $æ_2$ is smooth but it undergoes more distortion in the wider frequency band, and at a certain incident field intensity this resonance can overlap the first one (Fig. 8b). Evidently that in this case the transmission coefficient has more than two stable states, i.e. the effect of multistability occurs.

## 1.3 All-Dielectric Planar Metamaterials

### 1.3.1 Trapped-Modes in All-Dielectric Arrays

Unfortunately huge energy dissipation which is inherent to metal in the infrared and visible parts of spectrum results in increasing ohmic losses in plasmonic metamaterials [22-25] and decreasing $Q$-factor of the trapped-mode resonance [5]. Moreover, the trapped-mode resonance completely degrades in metamaterials with low degree of the particle asymmetry. Thus, using all-dielectric low-loss structures which are capable to support the trapped-mode excitation in the infrared and visible ranges is extremely good idea [26].

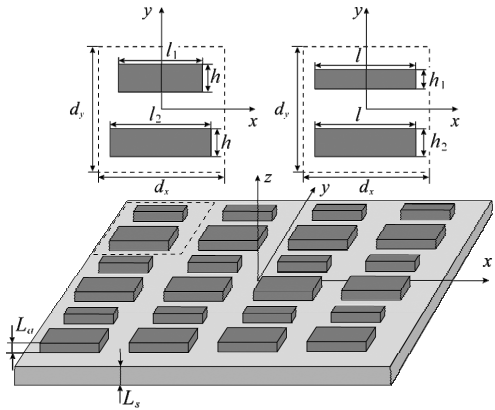

**Fig. 9.** Fragment of the all-dielectric metamaterial and its elementary unit cells.

Let us consider the plane wave diffraction on a double-periodic structure which consists of two dielectric elements within the periodic cell (Fig. 9). These dielectric elements create an electromagnetic environment similar to that one existed in the open dielectric resonators, and, unlike structures with metallic particles, a form of these dielectric elements does not entail a substantial increase of the resonant wavelength. Note that a material with high refractive index is required for designing the array of subwavelength elements to provide the resonant light interaction with the system. In particular, we propose to construct such elements in the form of two closely spaced parallelepipeds with different length or width similar to those ones which are shown in Fig. 9. In particular, in this figure the double-periodic array of dielectric bars placed on a silica substrate with thickness $L_s$ is presented. The unit cell of the array includes a pair of dielectric bars which have different length ( $l_1 \leq l_2$ , $h_1 = h_2 = h$ ) or width ( $h_1 \leq h_2$ , $l_1 = l_2 = l$ ) but are identical in thickness ($L_a$) and are made from the same material. The sizes of the square periodic cell are chosen to be identical $d_x = d_y = d$. The period cells of both structures are symmetric relative to the line drawn through the cell center parallel to the $y$-axis.

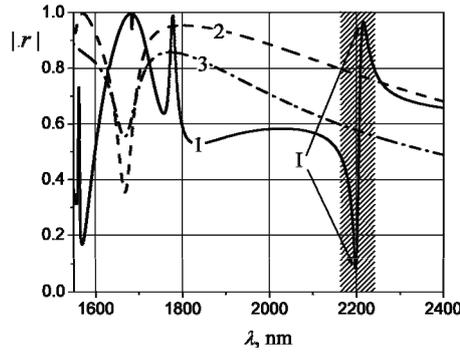
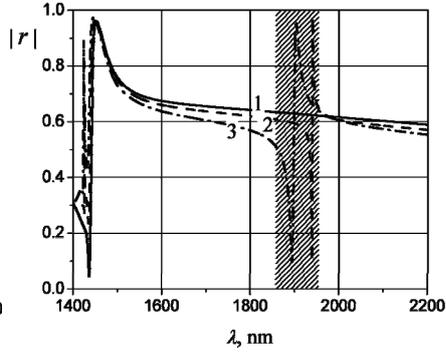

**Fig. 10.** Wavelength dependences of the reflection coefficient magnitude of the metamaterial with single dielectric bar (lines 2 and 3) and pair of dielectric bars (line 1) in periodic cell; $d = 975$ nm, $L_s = \infty$, $L_a = h = 195$ nm, $n_d = 5.5$. Line 1 corresponds to bars with sizes $l_1 = 780$ nm and $l_2 = 877$ nm, line 2 corresponds to bars with size $l = 877$ nm, and line 3 corresponds to bars with size $l = 780$ nm.

**Fig. 11.** Wavelength dependences of the reflection coefficient magnitude of the metamaterial with pair of germanium bars in periodic cell; $d = 975$ nm, $L_s = \infty$, $L_a = h = 195$ nm, $l_2 = 877$ nm, the germanium refractive index is $n_d = 4.12$ which is actual for wavelength 1900 nm. Line 1, 2 and 3 correspond to bars with sizes $l_1 = 877$ nm, $l_1 = 838$ nm, and $l_1 = 780$ nm, respectively.

We suppose a normal incidence of the linearly $x$-polarized plane waves on the structure. The resonant response of the array is studied in the near-infrared wavelength range from 1000 nm to 2500 nm. The substrate is assumed to be made from the synthetic fused silica whose refractive index is approximately 1.44 in the wavelength range under consideration [27]. Refractive index of the dielectric bars is $n_d$. The diffraction problem is solved numerically using the mapped PSTD method [28, 29].

The wavelength dependences of the reflection coefficient magnitude of the arrays with a single dielectric bar (lines 1 and 2) and a pair of dielectric bars (line 3) within the periodic cell are shown in Fig. 10. In this figure the arrows and Roman numeral I indicate the high-$Q$ resonance which appears in the array consisted of a pair of dielectric bars with different length. Each dielectric bar within the periodic cell interacts with light like a half-wavelength open dielectric resonator and the resulting field has antiphase distribution within the pair of these resonators, and, hence, this resonant regime can be referred to the trapped-mode excitation [6]. The main distinctive feature of the trapped-mode resonance in the two-element dielectric array is a great red shift of its wavelength compared to the resonant wavelengths of the corresponding single-element arrays (see Fig. 10). This feature of the trapped-mode resonant regime of the proposed all-dielectric metamaterial is quite important in view of its possible application in the infrared and visible parts of spectrum. First, the ratio of the array pitch to the wavelength may be decreased to design more homogeneous metamaterials. Second, as the resonant wavelength shifts up the field confinement increases since the radiation losses decrease. Remarkably, these behaviors are especially important when designing artificial nonlinear and gain artificial media.

The proposed array can be made of semiconductor in the wavelength range where semiconductor has a transparency window. In particular, the transparency windows of germanium and silicon lie between 1.9 $\mu$m and 16 $\mu$m, and 1.2 $\mu$m and 14 $\mu$m, respectively [30]. The semiconductor interacts with light as a good dielectric within these transparency windows. The typical value of the dissipation losses tangent of the mentioned semiconductors within these bands do not exceed $10^{-3}$. Also the semiconductor refractive index has extremely small variation of its value within the transparency windows. The germanium refractive index changes from 4.15 to 4.0, and silicon refractive index changes from 3.41 to 3.52, respectively.

The wavelength dependences of the reflection coefficient magnitude for the periodic array made from a pair of germanium bars in the periodic cell are shown in Fig. 11. One can see that both the $Q$-factor and the value of red shift of the trapped-mode resonance increase as the asymmetry of bars within the periodic cell decreases. The calculated $Q$-factors of the trapped-mode resonances are 203 and 1080 for the structures with germanium bars having geometrical parameters $l_1 = 780$ nm, $l_2 = 877$ nm and $l_1 = 838$ nm, $l_2 = 877$ nm, respectively (see lines 2 and 3 in Fig. 11). These values of $Q$-factor are ten orders of magnitude greater than those ones reached in the plasmonic metamaterials.

### 1.3.2 Saturation Effect in Active Metamaterial

We have proposed a simple design of an all-dielectric low-loss silicon-based planar metamaterial [31] which can produce an extremely sharp resonant reflection and transmission in the wavelength of about 1550 nm due to both low dissipation losses and the trapped-mode excitation. The *Q*-factor of the resonance exceeds in ten times the *Q*-factor of resonances in known plasmonic structures. The designed metamaterial is envisioned for aggregating with a pumped gain medium to achieve an enhancement of luminescence, and we report that in the designed metamaterial the essential enhancement of luminescence (more than 500 times) in a layer which consists of pumped quantum dots (QD) can be reached. This value significantly exceeds the known values of the luminescence enhancement in known plasmonic planar metamaterials [32].

The model of a gain nonlinear medium assumes the introducing negative frequency dependent conductivity in the form:

$$\sigma(\omega) = \frac{1}{1 + I/I_s} \cdot \frac{\sigma_0(1 + i\omega\tau)}{(1 + \omega_0^2\tau^2) + 2i\omega\tau - \omega^2\tau^2} \,, \tag{9}$$

where $\omega_0 = 1.26 \times 10^{15}\,\mathrm{s}^{-1}$ corresponds to the wavelength $\lambda_0 = 1550$ nm, $\tau = 4.85 \times 10^{-15}$ s, $\varepsilon_{QD} = 2.19$ corresponds to the refractive index $n_{QD} = 1.48$ of the non-pumped quantum dot laser medium, and $\sigma_0 = -500$ Sm/m corresponds to an emission factor $\mathrm{tg}\delta_\varepsilon = -0.021$ by analogy with a lossy factor of media. Small value of $\tau$ results in a wide-band QD spectral line and enables us to exclude from consideration the effects caused by displacement of metamaterial dissipation peak and maximum of exciton emission line of QDs. Let us notice that the pump level ($\sigma_0$) is in one order of magnitude less than it is needed in the case of plasmonic metamaterials because there is a low degree of losses in the all-dielectric array. The factor $(1 + I/I_s)^{-1}$ allows us to consider the effect of the luminescence enhancement of the gain saturation effect which is inherent to the active media. Here the parameter $I_s$ is proportional to the saturation intensity and displays the effect of inversion population reducing in the gain medium by simulated emission. It is proportional to the maximum of the internal field ($I$). The saturation factor $[(1 + I/I_s)^{-1}]$ is calculated separately for each point of the spatial grid which takes into account the heterogeneity of the QD layer. We should note that the small value of an emission factor results in independence of the QD refractive index on the saturation factor. Thus the effect of saturation versus the luminescence enhancement of QD layer hybridized with the all-dielectric metamaterial can be considered under this model.

The diffraction approach proposed in [31] is further used to calculate the luminescence enhancement in the QD layer hybridized with the all-dielectric metamaterial. This approach consists of evaluation of considered structure luminescence through the difference of the energy dissipation in the passive and active (pumped) structure. The dissipation energy is calculated from the solution of the corresponding diffraction problem for the separate plane wave. The luminescence enhancement equals to the ratio of the luminescence of the hybridized structure to the luminescence of the 210 nm homogeneous QD layer placed on 50 nm silica substrate. The wavelength dependences of the luminescence enhancement of the QD layer hybridized with the all-dielectric metamaterial for different values of the saturation intensity are shown in Fig. 12. The reducing of the luminescence enhancement with decreasing of the saturation intensity can be explained by exciting strong local field in the hybridized structure which results in the saturation factor decreasing. The distribution of the saturation factor in cross section ($z = -155$ nm) is depicted in Fig. 13. One can see the burning holes appearance in the distribution (see dark areas in Fig. 13). The energy of optical pumping within these holes is completely spent by simulated emission. The effect of gain saturation does not strongly affect on the photoluminescence in the system but it needs to be taken into account when designing optical amplifiers and lasing spacers.

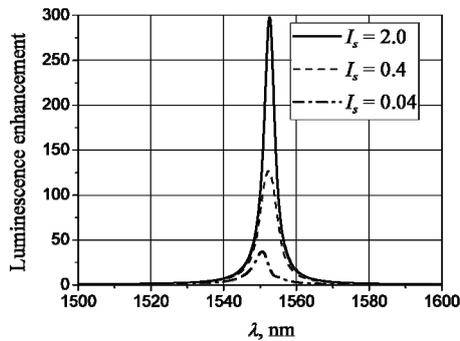
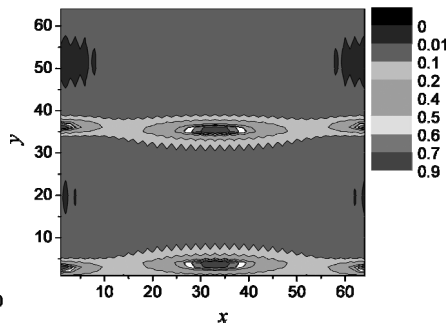

**Fig. 12.** The wavelength dependences of the luminescence enhancement (in a.u.) of the QD layer hybridized with the all-dielectric metamaterial for different value of $I_s$.

**Fig. 13.** The distribution of saturation factor in the cross section $z = -155$ nm; $\lambda = 1553$ nm, $I_s = 0.4$.

**Acknowledgements** This work was supported by the Ukrainian State Foundation for Basic Research, the Project no. Φ54.1/004, and National Academy of Sciences of Ukraine, Program 'Nanotechnologies and Nanomaterials', the Project no. 1.1.3.17.

# References

1. Rockstuhl C., Zentgraf T., Guo H., Liu N., Etrich C., Loa I., Syassen K., Kuhl J., Lederer F., Giessen H. Resonances of split-ring resonator metamaterials in the near infrared, *Appl. Phys. B*, **84**(1-2), 219-227 (2006).
2. Liu N., Guo H., Fu L., Schweizer H., Kaiser S., Giessen H. Electromagnetic resonances in single and double split-ring resonator metamaterials in the near infrared spectral region, *Phys. Stat. Sol. (b)*, **244**(4), 1251-1255 (2007).
3. Prosvirnin S., Zouhdi S., Resonances of closed modes in thin arrays of complex particles, in Advances in Electromagnetics of Complex Media and Metamaterials, edited by S. Zouhdi, M. Arsalane, 281-290 (Dordrecht: Kluwer Academic Publishers, 2003).

4.  Fedotov V.A., Rose M., Prosvirnin S.L., Papasimakis N., Zheludev N.I. Sharp trapped-mode resonances in planar metamaterials with a broken structural symmetry, *Phys. Rev. Lett.*, **99**(14), 147401 (2007).

5.  Khardikov V., Iarko E., Prosvirnin S., Trapping of light by metal arrays, *J. Opt.*, **12**(4), 045102 (2010).

6.  Samson Z.L., MacDonald K.F., DeAngelis F., Gholipour B., Knight K., Huang C.C., Di Fabrizio E., Hewak D.W., Zheludev N.I. Metamaterial electro-optic switch of nanoscale thickness, *Appl. Phys. Lett.*, **96**(14), 143105 (2010).

7.  Bergman D.J., Stockman M.I. Surface plasmon amplification by stimulated emission of radiation: Quantum generation of coherent surface plasmons in nanosystems, *Phys. Rev. Lett.*, **90**(2), 027402, (2003).

8.  Bergman D.J., Stockman M.I. Can we make a nanoscopic laser? *Laser Phys.*, **14**(3), 409–411, (2004).

9.  Zheludev N.I., Prosvirnin S.L., Papasimakis N., Fedotov V.A. Lasing spacer, *Nature Photonics*, **2**(6), 351-354 (2008).

10. Tuz V., Prosvirnin S., Kochetova L. Optical bistability involving planar metamaterials with broken structural symmetry, *Phys. Rev. B*, **82**(23), 233402 (2010).

11. Tuz V., Prosvirnin S. All-optical switching in metamaterial with high structural symmetry - Bistable response of nonlinear double-ring planar metamaterial, *Eur. Phys. J. Appl. Phys.*, **56**(3), 30401 (2011).

12. Tuz V., Butylkin V., Prosvirnin S. Enhancement of absorption bistability by trapping-light planar metamaterial, *J. Opt.*, **14**(4), 045102 (2012).

13. Dmitriev V., Prosvirnin S., Tuz V., Kawakatsu M. Electromagnetic controllable surfaces based on trapped-mode effect, *Adv. Electromagn.*, **1**(2), 89-95 (2012).

14. Gibbs H.M. Optical bistability: Controlling light with light (Orlando, FL: Academic, 1985).

15. Klein M.W., Enkrich C., Wegener M., Linden S., Second-harmonic generation from magnetic metamaterials, *Science*, **313**(5786), 502-504 (2006).

16. Prosvirnin S.L., Zouhdi S. Multi-layered arrays of conducting strips: Switchable photonic band gap structures, *AEÜ Int. J. Electron. Commun*, **55**(4), 260-265 (2001)

17. Prosvirnin S.L., Papasimakis N., Fedotov V., Zouhdi S., Zheludev N. Trapped-mode resonances in planar metamaterials with high structural symmetry, in Metamaterials and Plasmonics: Fundamentals, Modelling, Applications, edited by S. Zouhdi et al., 201-208 (The Netherlands: Springer, 2009).

18. Tarapov S.I., Machekhin Yu.P., Zamkovoy A.S. Magnetic resonance for optoelectronic materials investigating. (Kharkov: Collegium, 2008).

19. Fedotov V.A., Mladyonov P.L., Prosvirnin S.L., Zheludev N.I. Planar electromagnetic metamaterial with a fish scale structure, *Phys. Rev. E*, **72**(5), 056613 (2005).

20. Papasimakis N., Fedotov V.A., Zheludev N.I., Prosvirnin S.L. Metamaterial analog of electromagnetically induced transparency, *Phys. Rev. Lett.*, **101**(25), 253903, (2008).

21. Mladyonov P.L., Prosvirnin S.L. Wave diffraction by double-periodic gratings of continuous curvilinear metal strips placed on both sides of a dielectric layer, *Radio Physics and Radio Astronomy*, **1**(4), 309-320, (2010).

22. Jansen C., Al-Naib I., Born N., Koch M. Terahertz metasurfaces with high Q-factors, *Appl. Phys. Lett.*, **98**(5), 051109 (2011).

23. Al-Naib I., Singh R., Rockstuhl C., Lederer F., Delprat S., Rocheleau D., Chaker M., Ozaki T., Morandotti R. *Appl. Phys. Lett.*, **101**(7), 071108, (2012).

24. Zhang S, Genov D.A., Wang Y, Liu M, Zhang X. Plasmon-induced transparency in metamaterials, *Phys. Rev. Lett.*, **101**(4), 047401, (2008).

25. Dong Z.-G., Liu H., Xu M.-X., Li T., Wang S.-M., Zhu S.-N., Zhang X. Plasmonically induced transparent magnetic resonance in a metallic metamaterial composed of asymmetric double bars, *Opt Express*, **18**(17), 18229-18234, (2010).

26. Khardikov V., Iarko E., Prosvirnin S. A giant red shift and enhancement of the light confinement in a planar array of dielectric bars, *J. Opt.*, **14**(3), 035103 (2012).
27. Malitson I.H. Interspecimen comparison of the refractive index of fused silica, *J. Opt. Soc. Am.*, **55**(11), 1205-1208, (1965).
28. Xian G., Mirotznik M.S., Shi S., Prather D.W. Applying a mapped pseudospectral time-domain method in simulating diffractive optical elements, *J. Opt. Soc. Am. A*, **21**(5), 777-785, (2004).
29. Khardikov V.V., Iarko E.O., Prosvirnin S.L. Using transmission matrix and pseudospectral time-domain method to study of light diffraction on planar periodic structures, Radiophysics and Radioastronomy, **13**(2), 146-158, (2008).
30. Liu H.H. Refractive index of silicon and germanium and its wavelength and temperature derivatives, *J. Phys. Chem. Ref. Data*, **9**(3), 561-658, (1980).
31. Khardikov V.V., Prosvirnin S.L. Enhancement of quantum dot luminescence in all-dielectric metamaterial, *arXiv:1210.4146 [physics.optics]*, October 2012.
32. Tanaka K., Plum E., Ou J.Y., Uchino T., Zheludev N.I. Multifold enhancement of quantum dot luminescence in plasmonic metamaterials, *Phys. Rev. Lett.*, **105**(22), 227403, (2010).